\documentclass[%
 reprint,
 amsmath,amssymb,
 aip,jcp,
floatfix,
]{revtex4-1}

\usepackage{graphicx}
\usepackage{dcolumn}
\usepackage{bm}
\usepackage{marginnote}

\usepackage{hyperref,color,pgfplots,siunitx,multirow,braket}
\usetikzlibrary{plotmarks}
\pgfplotsset{compat=1.13} 

\usepackage{xargs}
\usepackage[colorinlistoftodos,prependcaption,textsize=footnotesize]{todonotes}
 \newcommandx{\unsure}[2][1=]{\todo[linecolor=red,backgroundcolor=red!25,bordercolor=red,#1]{#2}\ }
 \newcommandx{\change}[2][1=]{\todo[linecolor=blue,backgroundcolor=blue!25,bordercolor=blue,#1]{#2}\ }
 \newcommandx{\info}[2][1=]{\todo[linecolor=yellow,backgroundcolor=yellow!25,bordercolor=yellow,#1]{#2}\ }
 \newcommandx{\add}[2][1=]{\todo[linecolor=green,backgroundcolor=green!25,bordercolor=green,#1]{#2}\ }

\usepackage[normalem]{ulem}

\makeatletter
\ams@newcommand{\iiiiint}{\DOTSI\protect\MultiIntegral{5}}
\renewcommand{\MultiIntegral}[1]{%
  \edef\ints@c{\noexpand\intop
    \ifnum#1=\z@\noexpand\intdots@\else\noexpand\intkern@\fi
    \ifnum#1>\tw@\noexpand\intop\noexpand\intkern@\fi
    \ifnum#1>\thr@@\noexpand\intop\noexpand\intkern@\fi
    \ifnum#1>4 \noexpand\intop\noexpand\intkern@\fi 
    \noexpand\intop
    \noexpand\ilimits@
  }%
  \futurelet\@let@token\ints@a
}
\makeatother

\usepackage{upgreek}
\usepackage{caption}
\usepackage{subcaption}
\usepackage{tikz}
\usetikzlibrary{3d} 
\usepackage{float}

\begin{document}



\title{Indirect Communication Between Non-Markovian Baths.}

\author{Ben S. Humphries}
\author{Dale Green}
\affiliation{Faculty of Science, University of East Anglia, Norwich Research Park, Norwich, NR4 7TJ, UK}
\author{Garth A. Jones}
\email{garth.jones@uea.ac.uk}
\affiliation{Faculty of Science, University of East Anglia, Norwich Research Park, Norwich, NR4 7TJ, UK}

\begin{abstract}

In this work we develop a model for an undamped vibration in the presence of an overdamped bath. This two-bath model involves a new derivation of the hierarchical equations of motion (HEOM) for an overdamped Lorentz-Drude (LD) environment that is summed together with an undamped oscillator (UO) bath, termed LDUO-HEOM. We show that information transfer occurs between the two baths,  even in the absence of a direct coupling between the baths. This bath-system-bath, mediated information transfer leads to intricate non-Markovian dynamics. The model is analysed using expectation values of the bath coordinates and  generates 2D electronic spectra that are in qualitative agreement with single-bath models. Furthermore, the model eliminates the additional superfluous damping introduced by the finite spectral width of the underdamped bath in our previous two-bath underdamped-overdamped, `bath vibration model' [\textit{J. Chem. Phys.} 156, 084103 (2022)].

\end{abstract}

\maketitle

\section{Introduction}

Non-Markovianity in an open quantum system is characterized by the recurrence of information from the bath, back into the system. It has been hypothesized that non-Markovianity could be exploited as a quantum resource \cite{kuroiwa2021asymptotically, berk2023extracting}. An interesting question is: how does information dynamics behave in open systems with multiple baths~\cite{KoyanagiTanimura2024}? This question is relevant to a number of applications including quantum heat engines and quantum refrigerators. 
~\cite{KoyanagiTanimura2024,KoyanagiTanimura2024b,Pleasance2024,Friedman2018,Altintas2025,KatoTanimura2016,KilgourSegal2018,SongShi2017}

When considering this question, the construction of the open quantum system, and nature of the baths directly affects the information dynamics. In this work we consider a two bath open system comprising an undamped bath and an overdamped bath, where the baths are combined via a direct sum. As we will show, information exchange between the baths does occur despite the baths not being mixed via a tensor product. In essence, the system acts as an agent for transferring information between the baths via what we refer to as a bath-system-bath coupling.

In this work we use hierarchical equations of motion (HEOM) which are an important tool for modelling non-Markovian processes in a wide variety of applications including, but not restricted to, energy transduction, quantum information and quantum thermodynamics. While HEOM is incredibility powerful in its application and versatility, new equations must be derived from first principles for each new formulation. In recent years, there have been a number of new forms of HEOM~\cite{Tanimura2020c} derived for specific cases including arbitrary spectral density (ASD) HEOM~\cite{Tanimura1990}, dissipation equations of motion (DEOM)~\cite{Yan2016a}, and generalisations of HEOM~\cite{Ikeda2020,Ding2012,Wu2018}. These new derivations introduce additional flexibility into the decomposition of the Matsubara basis and hence account for different environmental structures.  Further, the definition of the system-bath boundary has been demonstrated to have a profound impact on the dynamics of the OQS model ~\cite{Seibt2020, Humphries2024, Iles-Smith2014, Maguire2019, McConnell2019}. This current work was motivated by some of our previous work where we considered system boundary placement by examining two different OQS models applied to a molecular system, where a vibrational mode is coupled to the electronic degrees of freedom. The models are called the Hamiltonian vibration model (HVM) and the bath vibration model (BVM). ~\cite{Humphries2022} In the HVM the intramolecular vibration is defined explicitly in the Hamiltonian and in the BVM the same vibration is removed from the Hamiltonian and subsumed into the bath via a canonical transformation.  In practice, the BVM was constructed by combining an underdamped bath \textit{added} to an overdamped bath, where the former describes the molecular vibrations and the latter the rest of the environment. In principle the HVM and BVM should be mathematically equivalent however we showed this was not the case in practice. This is because the in the case of the BVM, the additional bath gives rise to additional spectral broadening, that leads to superfluous damping that is not consistent with HVM dynamics and its resulting spectra. Our new derivation removes the spurious damping, without the computational expense of the ASD-HEOM. This is achieved through construction of a HEOM with two spectral density components, one overdamped (with a Lorentz-Drude form), and the other a completely undamped mode describing the pure intramolecular vibration.
The derivation follows well established methods of Tanimura and co-workers~\cite{Tanimura2012,Tanimura2006,Tanimura2014}. As discussed in Humphries et. al.~\cite{Humphries2022}, the BVM results in intrinsic canonical damping from the underdamped mode, that we argued is superfluous when added to a second overdamped bath, which already describes the noisy environment. This originates from the canonical transform which carries the underdamped vibration into the environmental degrees of freedom. By creating a hierarchy which contains an \textit{undamped}, rather than underdamped, contribution we show that we can remove this artifact. 

In Section II we outline key differences between one- and two-bath models, including a brief overview of some relevant, recent studies. In Section III we give a detailed account of the derivation of our LDUO model. In Section IV we briefly detail the parameters used in the simulations employing the LDUO model. In Section V we present our results which includes and analysis of the resulting 2D electronic spectra and a detailed analysis bof that bath dynamics through expectation values of collective bath coordinates, before conclusions in Section VI.

\section{Background and Approach}

Undamped hierarchies have been constructed in earlier works \cite{LiuZhuBaiShi2014,YanXingShi2020}, where the stability of these models have been dramatically improved by some researchers \cite{Dunn2019}. Our work is a novel extension of these models through the addition of a second overdamped (Lorentz-Drude) bath to the undamped bath. This construction differs from previous works on canonically transformed, \textit{underdamped}, equations because of how the environment is constructed. In works by Garg \cite{Garg1985}, Lai \cite{Lai2021}, Green \cite{Green2019}, and Tanimura \cite{Tanimura2020a, Tanimura2012}, a \textit{pure} system vibrational mode is extracted from the Hamiltonian and placed via the canonical transform into a Markovian bath. This is shown schematically in Figure 1(a), where superimposing the pure harmonic frequency (described by a delta function) with a Markovian environment (described by white noise) results in an underdamped (i.e. Lorentzian) spectral density. In particular we note that this environment is described by a single bath. In such a model there is \textit{no} superfluous canonical damping as the environment is correctly described by the white noise component of the single spectral density. 

\begin{figure}\label{canonical_trans}
    \centering
    \includegraphics[width=\linewidth]{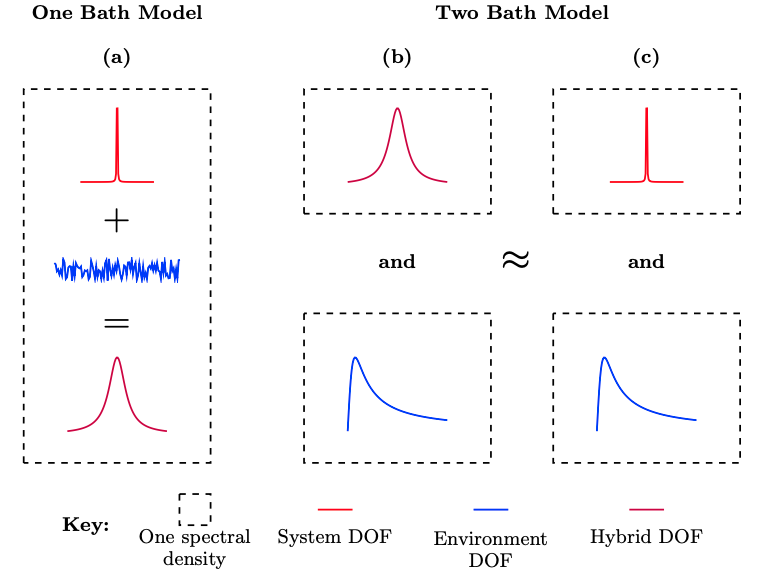}
    \caption{Schematic of one and two bath models.}
\end{figure}

Now we require the environment to be described by an overdamped spectral density, rather than purely white noise. To do this we adopt a \textit{two-bath approach} where the overdamped contributions are defined by a Lorentz-Drude spectral density. Importantly this means that the environment contributions will have some non-Markovian character, see Green et al. for more details \cite{Green2019}.  Previous studies canonically subsumed a system vibration into an overdamped environment by combining a Lorentzian with an overdamped spectral density, through a two-bath model. However, as we showed with the BVM, by doing this environmental damping is over-represented by doubling up on the environmental noise, see Figure 1(b), because there is noise in the Lorentzian as well as the LD spectral density. In 2022 we showed that a two-bath model in the limit of zero canonical damping is equivalent to a single-bath approach\cite{Humphries2022}. This motivated the current work where we now seek to develop a new model whereby the vibration remains in the bath but the white noise component is removed entirely. This is combined with an overdamped, Lorentz-Drude, spectral density as part of the HEOM derivation, thereby removing the superfluous damping, as seen in Figure 1(c).  

There are a number of motivational factors for doing this. Firstly it means that we can choose where to place the vibrational mode, the system Hamiltonian or the bath, which may have practical implications such as computing time. However, there is also significant theoretical interest, as it gives us insight into information dynamics between the bath and the system, which is an important area of study in open quantum systems. In particular in this work, we will investigate how information is exchanged between the two baths, via the quantum system. This is of general theoretical interest because non-Markovianity can be understood as a resource in quantum information theory. While a detailed analysis is outside the scope of the current work, this model has the potential to offer insight into how information might be fed between baths and back into a quantum system from its environment (recurrence).


\section{Model and derivation}

We consider a one-dimensional system coupled to a bath of $N$ harmonic oscillators. The system of interest is defined by an harmonic potential dependent on $\mathrm{q}$, the position operator of the system, with mass $m$, and vibrational frequency $\omega_{\mathrm{UO}}$. The oscillators in the bath have masses $\{m_j\}$, frequencies $\{\omega_j\}$ and coupling constants $\{c_j\}$. The general, total, Hamiltonian is, 
\begin{equation}
    \mathrm{H}_{\mathrm{tot}} = \mathrm{H}_{\mathrm{S}}+\mathrm{H}_{\mathrm{B}}+\mathrm{H}_{\mathrm{SB}},
\end{equation}
where,

\begin{equation}
    \mathrm{H}_{\mathrm{S}} = \ket{g}\,\mathrm{h}_g\,\bra{g}+\ket{e}\,\mathrm{h}_e\,\bra{e},
\end{equation}
with 
\begin{equation}
    \mathrm{h}_g = \frac{\mathrm{p}^2}{2m}+\frac{1}{2}m\omega_{\mathrm{UO}}^2\mathrm{q}^2,
\end{equation}
\begin{equation}
    \mathrm{h}_e = \hbar\omega_{eg}+\frac{\mathrm{p}^2}{2m}+\frac{1}{2}m\omega_{\mathrm{UO}}^2(\mathrm{q}-d)^2,
\end{equation}
\begin{equation}
    \mathrm{H}_{\mathrm{B}} = \sum_j\left[\frac{\mathrm{p}_j^2}{2m_j}+\frac{m_j\omega_j\mathrm{x}_j^2}{2}\right],
\end{equation}
\begin{equation}
    \mathrm{H}_{\mathrm{SB}} = -\sum_jc_j\mathrm{x}_j\mathrm{q},
\end{equation}
and $\mathrm{p},\mathrm{q},\{\mathrm{p}_j\},\{\mathrm{x}_j\}$ are the respective momentum and position operators for the system and the bath, and $d$ is the displacement of the excited  state potential energy surface. 

In the context of our derivation $d=0$, and the system Hamiltonian is simplified by canonically transforming the system frequency $\omega_{\mathrm{UO}}$ into the spectral density. This is what becomes the undamped oscillator mode, and hence is why the system vibration is named UO in the system Hamiltonian. This results in $\mathrm{H}_{\mathrm{S}}=\hbar\omega_{eg}\,\ket{e}\bra{e}$, and subsequently the interaction Hamiltonian depends on the bath coupling operator, rather than the system coordinate $\mathrm{H}_{\mathrm{SB}}=-\sum_j c_j\mathrm{x}_j\mathrm{B}$, with $\mathrm{B}= \ket{e}\bra{e}$.

Given this description of the system and bath we can construct the density matrix, starting from the system in the vibronic basis. For a set of states $\{\ket{\mathrm{q}_i,\mathrm{x}_i}\}$ with corresponding transition probabilities $\{P_i\}$ the density matrix is equivalent to the outer product,
\begin{equation}
    \uprho(\mathrm{q},\mathrm{x}) = \sum_i P_i\ket{\mathrm{q}_i,\mathrm{x}_i}\bra{\mathrm{q}_i,\mathrm{x}_i}.
\end{equation}
To derive the equations of motion we use the path integral formalism. Introducing the time propagation operators we can find the density matrix at an arbitrary, non-zero time, $t$ from an initial density matrix at time zero,
\begin{equation}
    \uprho_t(\mathrm{q},\mathrm{x}) = \exp\left(\frac{i\mathrm{H}(\mathrm{q},\mathrm{x})t}{\hbar}\right)\uprho_0(\mathrm{q},\mathrm{x})\exp\left(\frac{-i\mathrm{H}(\mathrm{q},\mathrm{x})t}{\hbar}\right).
\end{equation}
Next we apply the Born approximation so the system is initially in a factorisable state with respect to the thermally equilibrated bath, 
\begin{equation}
    \uprho_0(\mathrm{q},\mathrm{x}) = \uprho_{\mathrm{S}}(\mathrm{q})\uprho_{\mathrm{B}}(\mathrm{x}),
\end{equation}
which allows the reduced density matrix element to be expressed in path integral form, once the bath degrees of freedom have been traced out~\cite{Caldeira1983,Caldeira1981}:
\begin{multline}\label{dens_op}
    \uprho(\mathrm{q}_t,\mathrm{q}_t',t) = \iint{}\int^{\mathrm{q}_t}_{\mathrm{q}_0}\int^{\mathrm{q}_t'}_{\mathrm{q}_0'} \exp\Bigg{(}\frac{i\mathrm{S}_{\mathrm{S}}[\mathrm{q}_t]}{\hbar}\Bigg{)} \\ \times\exp\Bigg{(}-\frac{i\mathrm{S}_{\mathrm{S}}[\mathrm{q}_t']}{\hbar}\Bigg{)} \Bigg{(}\prod_n\mathcal{F}_n[\mathrm{q}_t,\mathrm{q}_t'] \Bigg{)}\uprho_{\mathrm{S}}(\mathrm{q}_0,\mathrm{q}_0',0) \\ \times\mathcal{D}[\mathrm{q}_t]\mathcal{D}[\mathrm{q}_t'] \ \text{d}\mathrm{q}_0 \ \text{d}\mathrm{q}_0',
\end{multline}
where $\int\mathcal{D}[\mathrm{q}_t]$ represents the functional integral. The action, denoted $\mathrm{S}_{\mathrm{S}}[\mathrm{q}_t;t]$, is associated with the corresponding system Hamiltonian, $\mathrm{H}_{\mathrm{S}}$. The bath effects are contained within the Feynman and Vernon influence functional for the $n$th mode with spectral density ~\cite{Caldeira1983,Feynman1963,Feynman1966}
\begin{multline}
    \mathcal{F}_n[\mathrm{q}_t,\mathrm{q}_t'] = \iiiiint{}\uprho_{\mathrm{B}}(\mathrm{x}_0,\mathrm{x}_0',0) \\ \times\exp\Bigg{(}\frac{i}{\hbar}\Big{[}\mathrm{S}_{\mathrm{B},n}[\mathrm{x}]-\mathrm{S}_{\mathrm{B},n}[\mathrm{x}'] +\mathrm{S}_{\mathrm{SB},n}[\mathrm{q}_t,\mathrm{x}]-\mathrm{S}_{\mathrm{SB},n}[\mathrm{q}_t',\mathrm{x}']\Big{]}\Bigg{)} \\ \times\mathcal{D}[\mathrm{x}]\mathcal{D}[\mathrm{x}']\text{d}\mathrm{x}_0\text{d}\mathrm{x}_0'\text{d}\mathrm{x}.
\end{multline}
The influence functional can be recast into a form that contains the kernels corresponding to fluctuation, $L_2^n(t)$,and dissipation, $iL_1^{n}(t)$~\cite{Feynman1963},
as in the original derivation by Tanimura and Kubo~\cite{Tanimura1989}. 
\begin{multline}
    \mathcal{F}_n=\exp\Bigg{(}-\frac{i}{\hbar}\int^{t}_{0}\int^{\tau}_{0}\mathrm{B}^{\times}(\mathrm{q}_t,\mathrm{q}_t';\tau) \\ \times\Bigg{[}iL_1^n(\tau-\tau')\mathrm{B}^{\circ}(\mathrm{q}_t,\mathrm{q}_t';\tau')+L_2^n(\tau-\tau')\mathrm{B}^{\times}(\mathrm{q}_t,\mathrm{q}_t';\tau')\Bigg{]} \\ \times\text{d}\tau' \ \text{d}\tau\Bigg{)},
\end{multline}
 with the general form of the bath coupling operator $\mathrm{B}$ in the vibronic basis obeying, $\mathrm{B}^{\times}(\tau)=\mathrm{B}^{\times}(\mathrm{q}_t;\tau)-\mathrm{B}^{\times}(\mathrm{q}'_t;\tau)$, and $\mathrm{B}^{\circ}(\tau)=\mathrm{B}^{\times}(\mathrm{q}_t;\tau)+\mathrm{B}^{\times}(\mathrm{q}'_t;\tau)$, respectively.
After canonical transformation from a vibronic system basis to an electronic system basis the continuous nuclear paths in system space, $q(\tau)$, become discrete system states $s(\tau)$ as a consequence of the removal of system vibrations. This leads to the simplification of the influence functional to
\begin{multline}
    \mathcal{F}_n=\exp\Bigg{(}-\frac{i}{\hbar}\int^{t}_{0}\int^{\tau}_{0}\mathrm{B}^{\times}(\mathrm{s}_t,\mathrm{s}_t';\tau) \\ \times\Bigg{[}iL_1^n(\tau-\tau')\mathrm{B}^{\circ}(\mathrm{s}_t,\mathrm{s}_t';\tau')+L_2^n(\tau-\tau')\mathrm{B}^{\times}(\mathrm{s}_t,\mathrm{s}_t';\tau')\Bigg{]} \\ \times\text{d}\tau' \ \text{d}\tau\Bigg{)},
\end{multline}
where the bath coupling operator $\mathrm{B}$ obeys the commutation and anti-commutation relations of the electronic system state, respectively, $\mathrm{B}^{\times}(\tau)=s(\tau)-s'(\tau)$, and $\mathrm{B}^{\circ}(\tau)=s(\tau)+s'(\tau)$.
\color{black} The kernels can be expressed by the spectral distribution as 
\begin{equation}
    L_2^n(t) = \int_0^{\infty}{}J_n(\omega)\cos(\omega t)\text{coth}\left(\frac{\beta\hbar\omega}{2}\right) \text{d}\omega,
\end{equation}
\begin{equation}
    iL_1^n(t) = -\int_0^{\infty}{}J_n(\omega)\sin(\omega t) \  \text{d}\omega.
\end{equation}
We split the environment into two major contributions: an overdamped bath (denoted $\mathrm{LD}$ for Lorentz-Drude), which is appropriate for Gaussian noise and an undamped oscillator ($\mathrm{UO}$) mode. The approach is similar to our BVM, where the undamped oscillator mode is subsumed canonically into the spectral density. This results in a form of the BVM, but in the limit of a zero linewidth vibration, where the canonical damping is zero. The spectral density can be written, 
\begin{equation}
    J(\omega)=J_{\mathrm{LD}}(\omega)+J_{\mathrm{UO}}(\omega),
\end{equation}
where 
\begin{equation*}
   J_{\mathrm{LD}}(\omega)=\frac{2\eta_{\mathrm{LD}}\gamma_{\mathrm{LD}}\omega_{0}^2\omega}{(\omega_{0}-\omega)^2+(\gamma_{\mathrm{LD}}\omega)^2},
\end{equation*}
which can be further simplified to the Lorentz-Drude form given that $\gamma_{\mathrm{LD}}\gg\omega_0$ such that $\Lambda_{\mathrm{LD}} = \omega_0^2/\gamma_{\mathrm{LD}}$,
\begin{equation}
    J_{\mathrm{LD}}(\omega)=\frac{2\eta_{\mathrm{LD}}\omega\Lambda_{\mathrm{LD}}}{\omega^2+\Lambda_{\mathrm{LD}}^2},
\end{equation}
and 
\begin{equation}
    J_{\mathrm{UO}} = \frac{1}{2}S_{\mathrm{UO}}^{\mathrm{HR}}\omega\left(\delta(\omega-\omega_{\mathrm{UO}})+\delta(\omega+\omega_{\mathrm{UO}})\right), 
\end{equation}
with $S_{\mathrm{UO}}^{\mathrm{HR}}=\lambda_{\mathrm{UO}}/\omega_{\mathrm{UO}}$. This form of the undamped part of spectral density matches that which is used in Seibt et al.~\cite{Seibt2018,Seibt2018a}, with the exception of the factor, $\omega_{\mathrm{UO}}$. This ensures that the spectral density has an amplitude proportional to the reorganisation energy, $\lambda_{\mathrm{UO}}$. We note that this is equivalent to applying a single delta function of the form $J(\omega)=S_{\mathrm{UO}}^{\mathrm{HR}}\omega_{\mathrm{UO}}\delta(\omega-\omega_{\mathrm{UO}})$.
The next step is to decompose the respective bath contributions, according to the Matsubara scheme, in order to explicitly incorporate time-dependent, but temperature-independent, Matsubara decomposition coefficients and frequencies~\cite{Seibt2018,Seibt2018a} into the correlation function. We perform this process for each component of the total spectral density piecewise. Starting with the Lorentz-Drude component we decompose the correlation function using complex contour integration~\cite{Humphries2024}, resulting in:
\begin{multline}
    L_{\mathrm{LD}}(t)=\eta_{\mathrm{LD}}\Lambda_{\mathrm{LD}}\left(\cot\left(\frac{\beta\hbar\Lambda_{\mathrm{LD}}}{2}\right)-i\right)e^{-\Lambda_{\mathrm{LD}} t} \\ +\sum_{n=1}^{\infty}\frac{2\eta_{\mathrm{LD}}\Lambda_{\mathrm{LD}}\nu_n}{\beta\hbar(\nu_n^2-\Lambda_{\mathrm{LD}}^2)}e^{-\nu_n t},
\end{multline}
which can be simplified by denoting,
\begin{equation}
    d_0=\eta_{\mathrm{LD}}\Lambda_{\mathrm{LD}}\left(\cot\left(\frac{\beta\hbar\Lambda_{\mathrm{LD}}}{2}\right)-i\right), \ \ \ \ \nu_0=\Lambda_{\mathrm{LD}},
\end{equation}
\begin{equation}
    d_n=\frac{2\eta_{\mathrm{LD}}\Lambda_{\mathrm{LD}}}{\beta\hbar}\left(\frac{\nu_n}{\nu_n^2-\Lambda_{\mathrm{LD}}^2}\right), \ \ \ \ \nu_n=\frac{2n\pi}{\beta\hbar},
\end{equation}
such that
\begin{equation}
    L_{\mathrm{LD}}(t)=\sum_{n=0}^{\infty}d_ne^{-{\nu_nt}}.
\end{equation}
Next we decompose the undamped oscillator mode component, proposed in reference \citenum{Seibt2020}, through the sifting property of the delta function resulting in:
\begin{multline}
    L_{\mathrm{UO}}(t) = \frac{S_{\mathrm{UO}}^{\mathrm{HR}}\omega_{\mathrm{UO}}}{2}\Bigg{[}\exp(-i\omega_{\mathrm{UO}}t)\Bigg{(}\text{coth}\Bigg{(}\frac{\beta\hbar\omega_{\mathrm{UO}}}{2}\Bigg{)} + 1 \Bigg{)} \\ +
    \exp(i\omega_{\mathrm{UO}}t)\Bigg{(}\text{coth}\Bigg{(}\frac{\beta\hbar\omega_{\mathrm{UO}}}{2}\Bigg{)} - 1\Bigg{)}\Bigg{]}.
\end{multline}
From this we have generated the Matsubara decomposition coefficients and frequencies for the second bath:
\begin{align}\label{ck}
    c_1 & = c_2^{*} = \frac{1}{2}S_{\mathrm{UO}}^{\mathrm{HR}}\omega_{\mathrm{UO}}\Bigg{(}\text{coth}\Bigg{(}\frac{\beta\hbar \ \omega_{\mathrm{UO}}}{2}\Bigg{)}+1\Bigg{)}, \\
    c_2 & = c_1^{*} = \frac{1}{2}S_{\mathrm{UO}}^{\mathrm{HR}}\omega_{\mathrm{UO}}\Bigg{(}\text{coth}\Bigg{(}\frac{\beta\hbar \ \omega_{\mathrm{UO}}}{2}\Bigg{)}-1\Bigg{)}, \\ 
    \gamma_1 & = \gamma_2^{*} = i\omega_{\mathrm{UO}}, \\
    \gamma_2 & = \gamma_1^{*} = -i\omega_{\mathrm{UO}}.
\end{align}
Additionally, the coordinates are moved into a coherent state basis transforming $\mathrm{a}_i^{\dagger}$ and $\mathrm{a}_i$, to $\mathrm{x}_j$ and $\mathrm{p}_j$ for the environment modes $\{j\}$. This representation uses
\begin{equation}
    \ket{\upphi}=\exp\Big{(}\sum_i\upphi_i\mathrm{a}^{\dagger}\Big{)}\ket{0},
\end{equation}
where $\ket{0}$ is the system vacuum sate, $\upphi_i$ are complex numbers, and $\upphi_i^{*}$ their complex conjugates such that
\begin{align}
    \mathrm{a}_i\ket{\upphi} & = \upphi_i\ket{\upphi}, \\
    \bra{\upphi}\mathrm{a}^{\dagger}_i & = \bra{\upphi}\upphi_i^{*}.
\end{align}
This is a movement from trajectories in physical space to trajectories of coherent states: $\mathrm{s}$, $\mathrm{x}$, $\mathrm{s}^{'}$ and $\mathrm{x}^{'}$ to $\mathrm{Q}_t=(\upphi^{*}(\tau),\upphi(\tau))$ and $\mathrm{Q}'_t=(\upphi'^{*}(\tau),\upphi'(\tau))$. 
Given these decompositions, and in a manner equivalent to the process in Ishizaki Tanimura~\cite{Ishizaki2005}, we construct the total influence function as:
\begin{equation}
\mathcal{F}=\prod_n\mathcal{F}_n = \mathcal{F}_{\mathrm{LD}}\times\mathcal{F}_{\mathrm{UO}},
\end{equation}
\begin{multline}
    \mathcal{F}_{\mathrm{UO}}=\exp\Bigg{(}-\frac{1}{\hbar}\int^{t}_{0}\int^{\tau}_{0}\sum_{k}\mathrm{B}_{k}^{\times}(\mathrm{Q}_t,\mathrm{Q}_t';\tau)\times \\ \exp(-\gamma_{k}(\tau-\tau'))\mathrm{\Theta}_k(\mathrm{Q}_t,\mathrm{Q}_t';\tau') \ \text{d}\tau' \ \text{d}{\tau}\Bigg{)}.
\end{multline}
where,
\begin{multline}
    \mathrm{\Theta}_k = \frac{1}{2}\Bigg{[}(c_k-c_k^{*})\mathrm{B}_{k}^{\circ}(\mathrm{Q}_t,\mathrm{Q}_t';\tau')+(c_k+c_k^{*})\mathrm{B}_{k}^{\times}(\mathrm{Q}_t,\mathrm{Q}_t';\tau')\Bigg{]},
\end{multline}
and 
\begin{multline}
    \mathcal{F}_{\mathrm{LD}} = \exp\Bigg{(}-\frac{1}{\hbar}\int^{t}_{0}\int^{\tau}_{0} \mathrm{B}^{\times}(\mathrm{Q}_t,\mathrm{Q}_t';\tau)\mathrm{\vartheta}(\mathrm{Q}_t,\mathrm{Q}_t';\tau')\Lambda_{\mathrm{LD}} \\ \times\exp(-\Lambda_{\mathrm{LD}}(\tau-\tau')) \ \text{d}{\tau'} \ \text{d}{\tau}\Bigg{)} \times
    \prod_{n=1}^{\infty}\exp\Bigg{(}-\frac{1}{\hbar}\int^{t}_{0}\int^{\tau}_{0} \\ \mathrm{B}^{\times}(\mathrm{Q}_t,\mathrm{Q}_t';\tau)\mathrm{\Psi}_n(\mathrm{Q}_t,\mathrm{Q}_t';\tau')\nu_n\exp(-\nu_n(\tau-\tau')) \ \text{d}{\tau'} \ \text{d}{\tau}\Bigg{)}.
\end{multline}
where,
\begin{multline}
    \mathrm{\vartheta}= \eta_{\mathrm{LD}}\Bigg{[}\text{cot}\Bigg{(}\frac{\beta\hbar\Lambda_{\mathrm{LD}}}{2}\Bigg{)}\mathrm{B}^{\times}(\mathrm{Q}_t,\mathrm{Q}_t';\tau')-i\mathrm{B}^{\circ}(\mathrm{Q}_t,\mathrm{Q}_t';\tau')\Bigg{]}, 
\end{multline}
\begin{equation}
    \mathrm{\Psi}_n = \sum_{n=1}^{\infty}\frac{2\eta_{\mathrm{LD}}\Lambda_{\mathrm{LD}}\nu_n}{\beta\hbar(\nu_n^2-\Lambda_{\mathrm{LD}}^2)}\mathrm{B}^{\times}(\mathrm{Q}_t,\mathrm{Q}_t';\tau').
\end{equation}
For a value of $K$, which satisfies $\nu_K=2\pi K/\beta\hbar\gg\omega_0$, where $\omega_0$ is the fundamental frequency of the system, then $\nu_n\exp(-\nu_n(\tau-\tau'))\approx\delta(\tau-\tau') \text{ for } n\geq K+1$. This simplifies the influence functional, equation (\ref{matrix_ele}).

We introduce the auxiliary operator,  $\uprho^{(m, l_k)}_{j_1\ldots j_{K}}$, by its matrix element\color{black}, in equation (\ref{ADO_matrix_element}),\color{black} as~\cite{Ishizaki2005}

\begin{widetext}
\begin{multline}\label{matrix_ele}
    \mathcal{F} \approx \exp\Bigg{(}-\int^{t}_{0}\int^{\tau}_{0}\sum_{k}\mathrm{B}_{k}^{\times}(\mathrm{Q}_t,\mathrm{Q}_t';\tau)\exp(-\gamma_{k}(\tau-\tau'))\mathrm{\Theta}_k(\mathrm{Q}_t,\mathrm{Q}_t';\tau') \ \text{d}{\tau'} \ \text{d}{\tau}\Bigg{)} \times \\ \exp\Bigg{(}-\int^{t}_{0}\mathrm{B}^{\times}(\mathrm{Q}_t,\mathrm{Q}_t';\tau)\exp(-\Lambda_{\mathrm{LD}}\tau)  \Bigg{[}-\int^{\tau}_{0}\Lambda_{\mathrm{LD}}\mathrm{\vartheta}(\mathrm{Q}_t,\mathrm{Q}_t';\tau')\exp(\Lambda_{\mathrm{LD}}\tau') \ \text{d}{\tau'}\Bigg{]} \ \text{d}{\tau}\Bigg{)}\times \\ \prod_{n=1}^{K} \exp\Bigg{(}-\int^{t}_{0}\mathrm{B}^{\times}(\mathrm{Q}_t,\mathrm{Q}_t';\tau)\exp(-\nu_n\tau)\Bigg{[}-\int^{\tau}_{0}\nu_n\mathrm{\Psi}_n(\mathrm{Q}_t,\mathrm{Q}_t';\tau')\exp(\nu_n\tau') \ \text{d}{\tau'}\Bigg{]} \ \text{d}{\tau} \Bigg{)}\times \\ \prod_{n=K+1}^{\infty}\exp\Bigg{(}\int_{0}^{t}\mathrm{B}^{\times}(\mathrm{Q}_t,\mathrm{Q}_t';\tau)\mathrm{\Psi}_n(\mathrm{Q}_t,\mathrm{Q}_t';\tau) \ \text{d}{\tau}\Bigg{)}.
\end{multline}

    \begin{multline} \label{ADO_matrix_element}
    \rho^{(m, l_k)}_{j_1\ldots j_{K}}(\mathrm{Q}_t,\mathrm{Q}_t';t)= \int_{\mathrm{Q}_t(t_0)}^{\mathrm{Q}_t(t)}\int_{\mathrm{Q}_t'(t_0)}^{\mathrm{Q}_t'(t)} \exp\Bigg{(}\frac{iS_{\mathrm{S}}[\mathrm{Q}_t,\mathrm{Q}_t']}{\hbar}\Bigg{)}\mathcal{F}\exp\Bigg{(}\frac{-iS_{\mathrm{S}}[\mathrm{Q}_t,\mathrm{Q}_t']}{\hbar}\Bigg{)} \uprho(\mathrm{Q}_{t_0},\mathrm{Q}_{t_0}';t_0) \times \\ 
    \prod_{k}\Bigg{\{} \int^{t}_{0}\exp(-\gamma_{k}(t-\tau'))\mathrm{\Theta}_k(\mathrm{Q}_t,\mathrm{Q}_t';\tau') \ \text{d}{\tau'}\Bigg{\}}^{l_k} \Bigg{\{} \exp(-\Lambda_{\mathrm{LD}}t)\Bigg{[}-\int_{0}^{t}\Lambda_{\mathrm{LD}}\mathrm{\vartheta}(\mathrm{Q}_t,\mathrm{Q}_t';\tau')\exp(\Lambda_{\mathrm{LD}}\tau') \ \text{d}{\tau'}\Bigg{]}\Bigg{\}}^{m} \times \\ \prod_{n=1}^{K}\Bigg{\{} \exp(-\nu_n t)\Bigg{[}-\int_{0}^{t}\nu_n\mathrm{\Psi}_n(\mathrm{Q}_t,\mathrm{Q}_t';\tau')\exp(\nu_n\tau') \ \text{d}{\tau'}\Bigg{]}\Bigg{\}}^{j_n} \mathcal{D}[\mathrm{Q}_t] \ \mathcal{D}[\mathrm{Q}_t'].
\end{multline}

\end{widetext}

for non-negative integers $l_k,m,j_1,\ldots,j_K$. 
Differentiating equation (\ref{ADO_matrix_element}) with respect to time, and then computing the path integrals, results in the following equations of motion~\cite{Humphries2024}: 
\begin{widetext}
\begin{multline}
    \frac{\partial}{\partial t}\uprho^{(m,l_k)}_{j_1\ldots j_{K}}=\Bigg{(}-\frac{i}{\hbar}\mathrm{H}_{\mathrm{S}}^{\times}-\sum_k(l_k\gamma_k+m\Lambda_{\mathrm{LD}})-\sum_{n=1}^{K}j_n\nu_n+\sum_{n=K+1}^{\infty}\mathrm{B}_{k}^{\times}\mathrm{\Psi}_n\Bigg{)}\uprho^{(m,l_k)}_{j_1\ldots j_{K}}- \\ \sum_{k}l_k\mathrm{\Theta}_k\uprho^{(m,l_k-1)}_{j_1\ldots j_{K}}-m\Lambda_{\mathrm{LD}}\mathrm{\vartheta}\uprho^{(m-1,l_k)}_{j_1\ldots j_{K}} -\sum_{n=1}^{K}j_n\nu_n\mathrm{\Psi}_n\uprho^{(m,l_k)}_{j_1\ldots j_{n-1}\ldots j_{K}} - \\ \Bigg{(}\mathrm{B}^{\times}\uprho^{(m+1,l_k)}_{j_1\ldots j_{K}}+\sum_k\mathrm{B}_{k}^{\times}\uprho^{(m,l_k+1)}_{j_1\ldots j_{K}}\Bigg{)} -\sum_{n=1}^{K}\mathrm{B}^{\times}\mathrm{\uprho}^{(m,l_k)}_{j_1\ldots j_{n+1}\ldots j_{K}}. 
\end{multline} 
\end{widetext}
Upon first inspection it may appear that there is an absent factor of $\gamma_k$ in the creation term from the $(l_k-1)$th Matsubara axis, however this is not the case. Based on the reduction criteria for the infinite Matsubara components, which for the overdamped contribution is: 
\begin{equation}
    \nu_{K} = \frac{2\pi K}{\beta\hbar}\gg \omega_0,
\end{equation}
we reduce to a delta function for a sufficient value of $K$. However, such a reduction cannot be performed for the undamped component. The undamped contribution introduces a pair of Matsubara decomposition coefficients and frequencies, as opposed to an infinite number, and as such a sufficient value of $K$ being chosen is unlikely. This lack of reduction motivates the current derivation and it results in factors of $(c_k\pm c_k^{*})$ in $\mathrm{\Theta}_k$, which (based on the form of $c_k$ in equation (\ref{ck})) accounts for the apparent missing factor of $\gamma_k$. 

\begin{figure*}[ht!]
    \centering
    \includegraphics[width=0.5\textwidth]{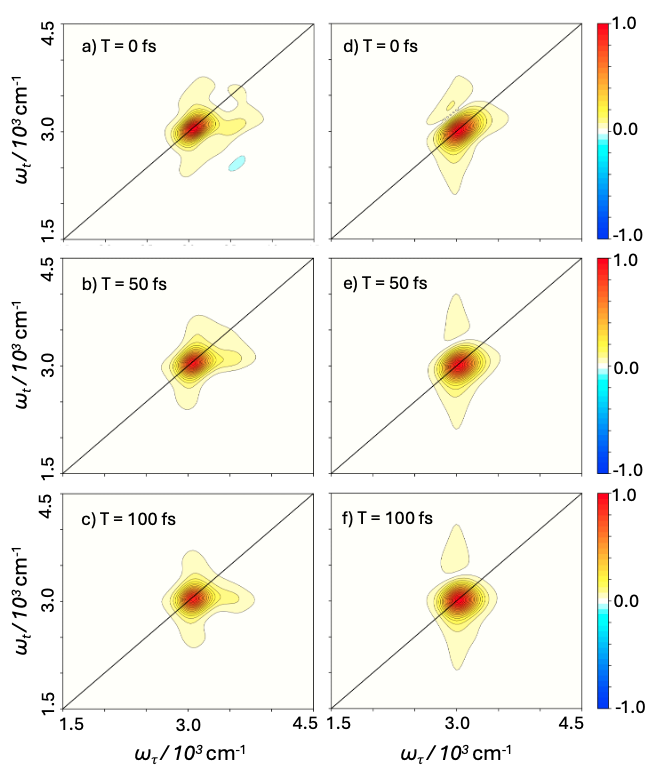}
    \caption{\label{fig:LDUO_2DES_limits} \color{black} Two-dimensional electronic spectra \color{black} (2DES) for the BVM system, column 1, and the LDUO-HEOM, column 2. Each column presents spectra for three population times, $T=0$, $50$, and $100$ fs.}
\end{figure*}

The condition, 
\begin{equation}
\sum_k(l_k\gamma_k+m\Lambda_{\mathrm{LD}})+\sum_{n=1}^{K}j_n\nu_n \gg \frac{\omega_0}{\mathrm{min}(\mathcal{I}(\nu_{k}),\mathcal{R}(\nu_n))},
\end{equation}
\begin{equation}
    \implies \ \ \Gamma_{\mathrm{max}}=10 \ \mathrm{max}\Big{(}\mathcal{I}(\gamma_{k})\Big{)}.
\end{equation}
 terminates the \color{black}over-\color{black} \textit{and} undamped axes. This leads to hierarchy volumes that are similar for both baths. 
 
 Subsequently, the infinite hierarchy can be truncated by the terminator~\cite{Dijkstra2017,Green2019}: 
\begin{equation}
\frac{\partial}{\partial t}\uprho^{(m,l_k)}_{j_1\ldots j_{K}}\approx\Bigg{(}-\frac{i}{\hbar}\mathrm{H}_{\mathrm{S}}^{\times}-\sum_kl_k\gamma_k + \sum_{n=K+1}^{\infty}\mathrm{B}^{\times}\mathrm{\Psi}_n\Bigg{)}\uprho^{(m,l_k)}_{j_1\ldots j_{K}}.
\end{equation}
Here, the phonon contributions from the system characteristic damping rate vanish as they are a purely real decay, whereas the purely imaginary oscillating components persist. This can be rationalised through the limit of infinite time, all contributions with an associated damping will vanish leaving only oscillatory components after the application of the Markovian criterion. This can be rewritten as~\cite{Tanaka2009}
\begin{equation}
\frac{\partial}{\partial t}\uprho^{(m,l_k)}_{j_1\ldots j_{K}}\approx\Bigg{(}-\frac{i}{\hbar}\mathrm{H}_{\mathrm{S}}^{\times}-i(l_0-l_1)\omega_{\mathrm{UO}} + \sum_{n=K+1}^{\infty}\mathrm{B}^{\times}\mathrm{\Psi}_n\Bigg{)}\uprho^{(m,l_k)}_{j_1\ldots j_{K}}.
\end{equation}
Since the undamped oscillator component contributes a pair of Matsubara decomposition frequencies and coefficients, we expect the number of hierarchy elements to be only slightly greater than that of an overdamped \color{black}HEOM\color{black}.

\section{Numerical simulations}

The objective of the above derivation is to produce a version of HEOM that will give accurate spectra for systems in which an undamped oscillation is coupled to an overdamped spectral density. This new model is designed to eliminate the additional damping that occurs from an underdamped spectral density with finite linewidth. We apply the LDUO-HEOM to nonlinear optical spectroscopy to test the effect of superfluous damping on quality of 2D spectra. Note that in generating the spectra, instead of using the squared macroscopic polarisation, in the case of an undamped mode it is necessary to sum the absorptive component with the $\pi^{\mathrm{c}}$ phase flipped dispersive component in order to preserve the spectral broadening.

To benchmark the new HEOM, a two-level system with a fundamental transition frequency of $\omega_{eg} = 3000 \ \mathrm{cm}^{-1}$ is used. Each electronic level possesses a set of $N$ vibrational states with vibrational frequency, $\omega_0 = 500 \ \mathrm{cm}^{-1}$. More details of the model can be found in Green et al. ~\cite{Green2019}


The bath parameters, used in these simulations are $\eta=(50, 50)$ cm$^{-1}$, and $\gamma=(100, 2500)$ cm$^{-1}$, for the BVM, the first value in parentheses corresponds to the overdamped environment and the second the canonically subsumed mode resulting in an underdamped bath, and $\eta=50$ cm$^{-1}$, $\Lambda=100$ cm$^{-1}$, and $\lambda_{\mathrm{UO}}=0.5$ cm$^{-1}$,  for the LDUO-HEOM. 

The left column of figure \ref{fig:LDUO_2DES_limits} shows the BVM case while the right shows the LDUO-HEOM case for the identical system. The two sets of spectra in figure \ref{fig:LDUO_2DES_limits} are qualitatively similar, both in terms of peak positions and broadening, demonstrating that the LDUO-HEOM is effective in modelling the BVM in the limit of vanishing canonical damping. Both the BVM and LDUO-HEOM spectra present an inhomogeneously broadened  fundamental peak which becomes more rounded for later population times, indicating spectral diffusion at later times. A notable consequence of the undamped mode is a vertical stretching (along the emission axis) of the peaks. This is expected and occurs because of the longer-lived oscillation of the polarisation that results from the undamped contribution to the total spectral density.

One of the most major benefits of this new approach, apart from the removal of the artificial damping, is the reduction in computational cost. In order to generate these spectra there is an equilibration step, an evolution, and calculation of 2D spectra. In the case of the BVM these steps take 1 hour 29 mins, 3 hours 34 mins, and 16 hours and 19 mins, respectively. However, in the LDUO-HEOM the same steps took $\sim 30$ seconds, $\sim 1$ minute, and $\sim 3$ minutes, respectively, on the same cluster. This means that the LDUO-HEOM took $0.56\%$ of the time the BVM did, or better, and represents an improvement of at least $99.4\%$. 

\section{Analysis of Bath Communication}

Having shown that the two-bath LDUO-HEOM  approach produces 2DES which are in qualitative agreement with analogous single-bath models, we now consider communication (i.e. information transfer) between the two baths. Firstly we note that fully coupled, product spectral densities and full generating functional approaches are yet to be applied to any two-bath models. When the baths are not directly coupled, the degree of implicit communication between the two constituent baths is dictated solely by their couplings to the quantum system. A previous study, Seibt et al. proposed models in which sub-components of the total system are explicitly coupled together via Hertzberg-Teller coupling through generating function approaches. \cite{Seibt2018} In our model, the system-bath boundary is not as clear cut because hybridisation between the degrees of freedom of the environment and the molecule is implicit in the construction of the LDUO-HEOM.  While the appearance of a vibronic progression in the 2DES, above, indicates that the electronic degrees of freedom from the system couple with vibrational information contained in the spectral density, this does not confirm whether the individual baths interact with each other in the absence of explicit bath-bath coupling. To quantify any explicit correlations between the baths, which would imply information transfer between them, we apply the bath coordinate expectation value approach proposed by Zhu et al \cite{Zhu2012,Shi2009}. The system-bath interaction term is taken as
\begin{equation}
    \mathrm{H}_{\mathrm{SB}}=-\sum_{j}c_j\mathrm{x}_j\mathrm{B} \equiv -\mathrm{F}\otimes \mathrm{B},
\end{equation}
where $\mathrm{F}=\sum_jc_j\mathrm{x}_j$ is a collective bath coordinate for the system. Zhu et al. showed that the expectation values of the collective bath coordinate, $\mathrm{F}$, can be tracked directly from the density matrix and its associated ADOs. The structure of the bath coordinate is such that it defines environmental polarisation effects in fermionic systems and energetic fluctuations due to excitation of the environment.  In both instances, this quantifies the degree of the system-bath interaction.  In HEOM methodologies the information exchanged between a system and bath is distributed throughout the ADOs and as such, the ADOs are intrinsic to the calculation of expectation operators of the bath coordinate. The hierarchy structure is reflects the properties of the OQS including the bath relaxation rate and system-bath coupling strength. 

Each Matsubara frequency corresponds to an independent axis within the hierarchy, and ADOs along this dimension contain quantum information in terms of integer multiples of the Matsubara frequency\cite{Humphries2024}. There have been a number of different ways to represent the ADO hierarchy geometrically presented over the years. \cite{Noack2018,Shi2009,Zhu2012,Tanimura2006,Tanimura2020a,Ishizaki2005}  In our analysis, we have found that a pyramidal structure is very intuitive for representing the flow of information through the ADOs, from the density operator in the apex. Figure \ref{tiers} shows a simple example of a hierarchical structure with three Matsubara dimensions $\nu_{k}, \, k\in\{1,2,3\}$. Tiers are defined as ADOs with a common sum of indices. For example, all Matsubara vectors with a total index sum of $1$ are first tier, and are shown on the blue surface within the hierarchy volume in fig. \ref{tiers}. Each tier exists as a surface within the Matsubara space which corresponds to a physical cutoff speed at which phonon modes are considered to be Markovian. Physically this corresponds to the extent to which information is able to extend through the hierarchy. At $\Gamma_{\mathrm{max}}$, or greater, ADOs contain phonon states which are faster than the non-Markovian cutoff and as such these ADOs will be outside the hierarchy volume in which information can return to the density matrix.

\begin{figure}
\begin{center}
\begin{tikzpicture}[scale=1.60]
    \begin{scope}[rotate around z=135, rotate around y=160, rotate around x=10]

    \draw[thick,->] (0,0,0) -- (3.2,0,0) node[right] {$\nu_0$}; 
    \draw[thick,->] (0,0,0) -- (0,3.05,0) node[below] {$\nu_1$}; 
    \draw[thick,->] (0,0,0) -- (0,0,-3.5) node[below] {$\nu_2$}; 
    
    \end{scope} 

    \begin{scope}[rotate around z=135, rotate around y=160, rotate around x=10]
    
    \foreach \x in {0,1,2,3} { 
        \foreach \y in {0,1,2,3} { 
            \foreach \z in {0,1,2,3} { 
                \pgfmathsetmacro{\sum}{\x+\y+\z}
                \pgfmathparse{\sum<4 ? 1 : 0} 
                \let\check=\pgfmathresult
                \ifnum\check=1 
                    \node[fill=black,circle,inner sep=0.5pt] at (\x, \y, -\z) {};
                    \node[anchor=south east,scale=0.6] at (\x, \y, -\z) {$(\x,\y,\z)$};
                \fi
            }
        }
    }

    \foreach \x in {0,1,2,3} {
        \foreach \y in {0,1,2,3} {
            \foreach \z in {0,1,2} { 
                \pgfmathsetmacro{\sum}{\x+\y+\z}
                \pgfmathparse{\sum<3 ? 1 : 0} 
                \let\check=\pgfmathresult
                \ifnum\check=1 
                    \draw[gray,thin] (\x,\y,-\z) -- (\x,\y,-\the\numexpr\z+1\relax);
                    \ifnum\x<3
                        \draw[gray,thin] (\x,\y,-\z) -- (\the\numexpr\x+1\relax,\y,-\z);
                    \fi
                    \ifnum\y<3
                        \draw[gray,thin] (\x,\y,-\z) -- (\x,\the\numexpr\y+1\relax,-\z);
                    \fi
                \fi
            }
        }
    }

\end{scope} 

    \begin{scope}[rotate around z=135, rotate around y=160, rotate around x=10]
        \fill[blue, opacity=0.3] 
        (0, 0, -1) coordinate (A) 
        (1, 0, 0) coordinate (B) 
        (0, 1, 0) coordinate (C) ;
        
        \filldraw[blue, opacity=0.3] (A) -- (B) -- (C) -- cycle;
    \end{scope} 

    \begin{scope}[rotate around z=135, rotate around y=160, rotate around x=10]
        \fill[blue, opacity=0.3] 
        (0, 0, -2) coordinate (A) 
        (2, 0, 0) coordinate (B) 
        (0, 2, 0) coordinate (C) ;
        
        \filldraw[red, opacity=0.3] (A) -- (B) -- (C) -- cycle;
    \end{scope} 

    \begin{scope}[rotate around z=135, rotate around y=160, rotate around x=10]
        \fill[blue, opacity=0.3] 
        (0, 0, -3) coordinate (A) 
        (3, 0, 0) coordinate (B) 
        (0, 3, 0) coordinate (C) ;
        
        \filldraw[green, opacity=0.3] (A) -- (B) -- (C) -- cycle;
    \end{scope} 

    \begin{scope}[shift={(-1.25,-2.5)}]
        \node[anchor=north] at (1.5, 0) {\textbf{Key:}};
        \node[fill=blue, opacity=0.3, minimum size=10pt, anchor=north] at (0, -0.5) {};
        \node[anchor=south] at (0.0, -1.0) {Tier 1 ADOs};
        \node[fill=red, opacity=0.3, minimum size=10pt, anchor=north] at (1.5, -0.5) {};
        \node[anchor=south] at (1.5, -1.0) {Tier 2 ADOs};
        \node[fill=green, opacity=0.3, minimum size=10pt, anchor=north] at (3, -0.5) {};
        \node[anchor=south] at (3, -1.0) {Tier 3 ADOs};
    \end{scope}
    
\end{tikzpicture}
\end{center}
\caption{A generalised hierarchy diagram showing the ADO Matsubara vectors for a system with three Matsubara dimensions, $\nu_k, \ \ k=\{1,2,3\}$. Tiers are shown as coloured surfaces within the closed pyramidal volume.} \label{tiers}
\end{figure}

Given the definition of tiers within the hierarchy, we calculate the bath coordinate expectation values at each tier in the hierarchy. To simplify the indexing, we define three vectors; a collective Matsubara frequency index, a coefficient index, and Matsubara index for the total HEOM as
\begin{equation}
    \nu^{\mathrm{LDUO}}_{\alpha}=[\gamma_k,\,\Lambda_{\mathrm{LD}},\,\nu_n],
\end{equation}
\begin{equation}
    e_{\alpha}=[c_k,\,d_0,\,d_n],
\end{equation}
\begin{equation}
    n_{\alpha}=[l_k,\,m,\,j_n].
\end{equation}
This means that the ADOs can be rewritten as:
\begin{equation}
    \uprho^{(m,l_k)}_{j_1\ldots j_{K}} \equiv \uprho_{n_{\alpha}},
\end{equation}
with the associated correlation function
\begin{equation}
    L_{\mathrm{tot}}(t)=\sum_{\alpha}e_{\alpha}e^{-\nu^{\mathrm{LDUO}}_{\alpha}t}.
\end{equation}
Consequently, the bath coordinate expectation value can be calculated as 
\begin{equation}
    \mathrm{X}^{(1)}= -\sum_{\mathrm{tier}\, 1}\uprho_{n_{\alpha}}, \ \ \ \sum_{\alpha}n_{\alpha}=1,
\end{equation}
\begin{equation}
    \mathrm{X}^{(2)}= \mathrm{A}_0+\underbrace{\sum_{\mathrm{tier}\, 2}\uprho_{n_{\alpha}}}_{n_{\alpha}=0, \, \mathrm{or} \, 2} +2\underbrace{\sum_{\mathrm{tier} \, 2}\uprho_{n_{\alpha}}}_{n_{\alpha}=0, \, \mathrm{or} \, 1}, \ \ \ \sum_{\alpha}n_{\alpha}=2,
\end{equation}
where
\begin{align}
    \mathrm{A}_0 &= \int_{0}^{\infty} \frac{J_{\mathrm{tot}}(\omega)}{\pi}\mathrm{coth}\Big{(}\frac{\beta\hbar\omega}{2}\Big{)} \,\mathrm{d}\omega, \\
    &= \sum_{\alpha} e_{\alpha}.
\end{align}
Higher orders of the bath coordinate expectation value can be generated from the recursion relation:
\begin{equation}
    \mathrm{X}^{(n+1)}=\sum_{i=0}^{n+1}L_i^{(n+1)}\sum_{\mathrm{tier} \, i}\frac{i\,!}{\prod_{\alpha}n_{\alpha}\,!}\uprho_{n_{\alpha}}, \ \ \ \sum_{\alpha}n_{\alpha}=i, 
\end{equation}
and 
\begin{equation}
    L^{(n+1)}_{i}=-L^{(n)}_{i-1}+n\mathrm{A}_0L_i^{(n-1)},
\end{equation}
with $L_0^0=1$.  In this work we are primarily interested in the first two orders of the bath coordinate expectation value because the focus is on comparing one- and two-component spectral densities. The interested reader should refer to Zhu et al.  \cite{Zhu2012,Shi2009} for further details on higher orders.  

In order to facilitate a direct comparison of the constituent modes we consider projections  the collective bath coordinate for each of the modes in turn. Geometrically, this corresponds to taking the sealed hierarchy volume from fig. \ref{tiers} and projecting it to a closed surface. In this way we are able to generate bath coordinate expectation values for the independent LD and UO components, as well as for the full LDUO. Figure \ref{matsubara_partial_trace} demonstrates the process of projecting out components of the collective bath coordinate so that only ADOs associated with the left face of the pyramid are extracted. 


\begin{figure*}[h!]
\begin{tikzpicture}
\node at (0,0) {
\begin{tikzpicture}[scale=1.6]
    \node at (-1.5,0) {$\mathrm{M}_{\mathrm{vec}}=$};
    \begin{scope}[rotate around z=135, rotate around y=160, rotate around x=10]

    \draw[thick,->] (0,0,0) -- (3.2,0,0) node[right] {$\nu_{\mathrm{Z}}$}; 
    \draw[thick,->] (0,0,0) -- (0,3.05,0) node[below] {$\nu_{\mathrm{X}}$}; 
    \draw[thick,->] (0,0,0) -- (0,0,-3.5) node[below] {$\nu_{\mathrm{Y}}$}; 
    
    \end{scope} 

    \begin{scope}[rotate around z=135, rotate around y=160, rotate around x=10]
    
    \foreach \x in {0,1,2,3} { 
        \foreach \y in {0,1,2,3} { 
            \foreach \z in {0,1,2,3} { 
                \pgfmathsetmacro{\sum}{\x+\y+\z}
                \pgfmathparse{\sum<4 ? 1 : 0} 
                \let\check=\pgfmathresult
                \ifnum\check=1 
                    \node[fill=black,circle,inner sep=0.5pt] at (\x, \y, -\z) {};
                    \node[anchor=south east,scale=0.6] at (\x, \y, -\z) {$(\x,\y,\z)$};
                \fi
            }
        }
    }

    \foreach \x in {0,1,2,3} {
        \foreach \y in {0,1,2,3} {
            \foreach \z in {0,1,2} { 
                \pgfmathsetmacro{\sum}{\x+\y+\z}
                \pgfmathparse{\sum<3 ? 1 : 0} 
                \let\check=\pgfmathresult
                \ifnum\check=1 
                    \draw[gray,thin] (\x,\y,-\z) -- (\x,\y,-\the\numexpr\z+1\relax);
                    \ifnum\x<3
                        \draw[gray,thin] (\x,\y,-\z) -- (\the\numexpr\x+1\relax,\y,-\z);
                    \fi
                    \ifnum\y<3
                        \draw[gray,thin] (\x,\y,-\z) -- (\x,\the\numexpr\y+1\relax,-\z);
                    \fi
                \fi
            }
        }
    }

\end{scope} 
    
\end{tikzpicture}
};
\node at (9.5,0) {
\begin{tikzpicture}[scale=1.6]
    \node at (-1.5,0) {$\mathrm{Pr}_{\mathrm{XY}}(\mathrm{M}_{\mathrm{vec}})=$};
    \begin{scope}[rotate around z=135, rotate around y=160, rotate around x=10]

    \draw[thick,->, opacity=0.3, color=white] (0,0,0) -- (3.2,0,0) node[right] {$\nu_{\mathrm{Z}}$}; 
    \draw[thick,->] (0,0,0) -- (0,3.05,0) node[below] {$\nu_{\mathrm{X}}$}; 
    \draw[thick,->] (0,0,0) -- (0,0,-3.5) node[below] {$\nu_{\mathrm{Y}}$}; 
    
    \end{scope} 

    \begin{scope}[rotate around z=135, rotate around y=160, rotate around x=10]
    
    \foreach \x in {0} { 
        \foreach \y in {0,1,2,3} { 
            \foreach \z in {0,1,2,3} { 
                \pgfmathsetmacro{\sum}{\x+\y+\z}
                \pgfmathparse{\sum<4 ? 1 : 0} 
                \let\check=\pgfmathresult
                \ifnum\check=1 
                    \node[fill=black,circle,inner sep=0.5pt] at (\x, \y, -\z) {};
                    \node[anchor=south east,scale=0.6] at (\x, \y, -\z) {$(\x,\y,\z)$};
                \fi
            }
        }
    }

    \foreach \x in {0} {
        \foreach \y in {0,1,2,3} {
            \foreach \z in {0,1,2} { 
                \pgfmathsetmacro{\sum}{\x+\y+\z}
                \pgfmathparse{\sum<3 ? 1 : 0} 
                \let\check=\pgfmathresult
                \ifnum\check=1 
                    \draw[gray,thin] (\x,\y,-\z) -- (\x,\y,-\the\numexpr\z+1\relax);
                    \ifnum\x<3
                        \draw[gray,thin] (\x,\y,-\z) -- (\the\numexpr\x\relax,\y,-\z);
                    \fi
                    \ifnum\y<3
                        \draw[gray,thin] (\x,\y,-\z) -- (\x,\the\numexpr\y+1\relax,-\z);
                    \fi
                \fi
            }
        }
    }

    \foreach \x in {0,1,2,3} {
        \foreach \y in {0,1,2,3} {
            \foreach \z in {0,1,2} { 
                \pgfmathsetmacro{\sum}{\x+\y+\z}
                \pgfmathparse{\sum<3 ? 1 : 0} 
                \let\check=\pgfmathresult
                \ifnum\check=1 
                    \draw[gray,thin, opacity=0.4] (\x,\y,-\z) -- (\x,\y,-\the\numexpr\z+1\relax);
                    \ifnum\x<3
                        \draw[gray,thin, opacity=0.4] (\x,\y,-\z) -- (\the\numexpr\x+1\relax,\y,-\z);
                    \fi
                    \ifnum\y<3
                        \draw[gray,thin, opacity=0.4] (\x,\y,-\z) -- (\x,\the\numexpr\y+1\relax,-\z);
                    \fi
                \fi
            }
        }
    }

\end{scope} 
    
\end{tikzpicture}
};
\end{tikzpicture}
\caption{A schematic of taking the projection onto the XY plane constituent of the collective bath mode resulting in the removal of the $\nu_{\mathrm{Z}}$ Matsubara dimension from a generalised, 3D, hierarchy.}\label{matsubara_partial_trace}
\end{figure*}




\begin{figure*}[h!]
    \centering
    \includegraphics[width=0.95\textwidth]{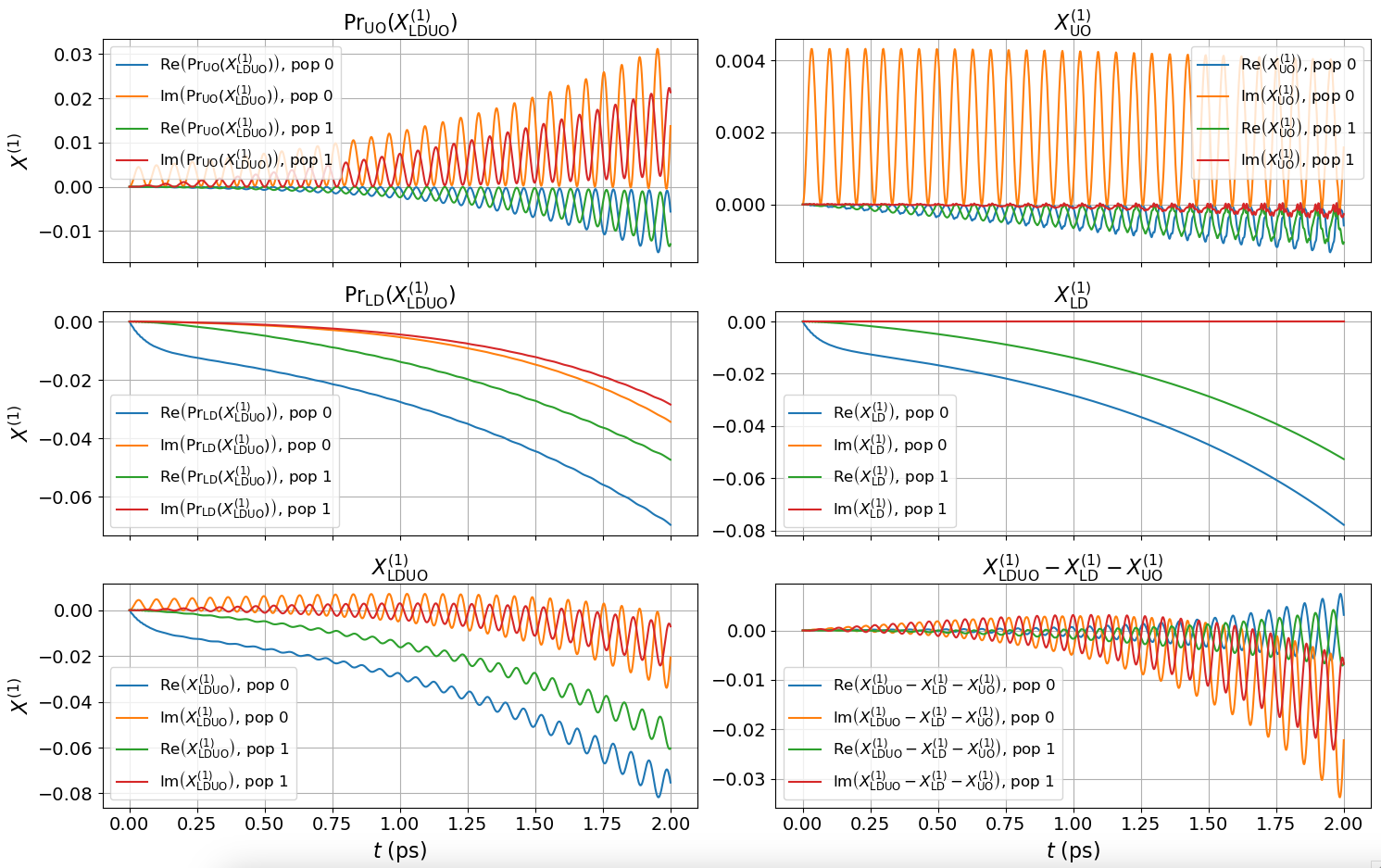}
    \caption{\label{fig:LDUO_diff_1} Tier 1 expectation values. Column 1, the bath coordinate of the full LDUO, projected onto the UO plane, projected onto the LD plane, or unaltered. Column 2, the bath coordinate of the UO, the LD, and the difference between the full bath coordinate and the sum of its parts.}
\end{figure*}

\begin{figure*}[h!]
    \centering
   \includegraphics[width=0.95\textwidth]{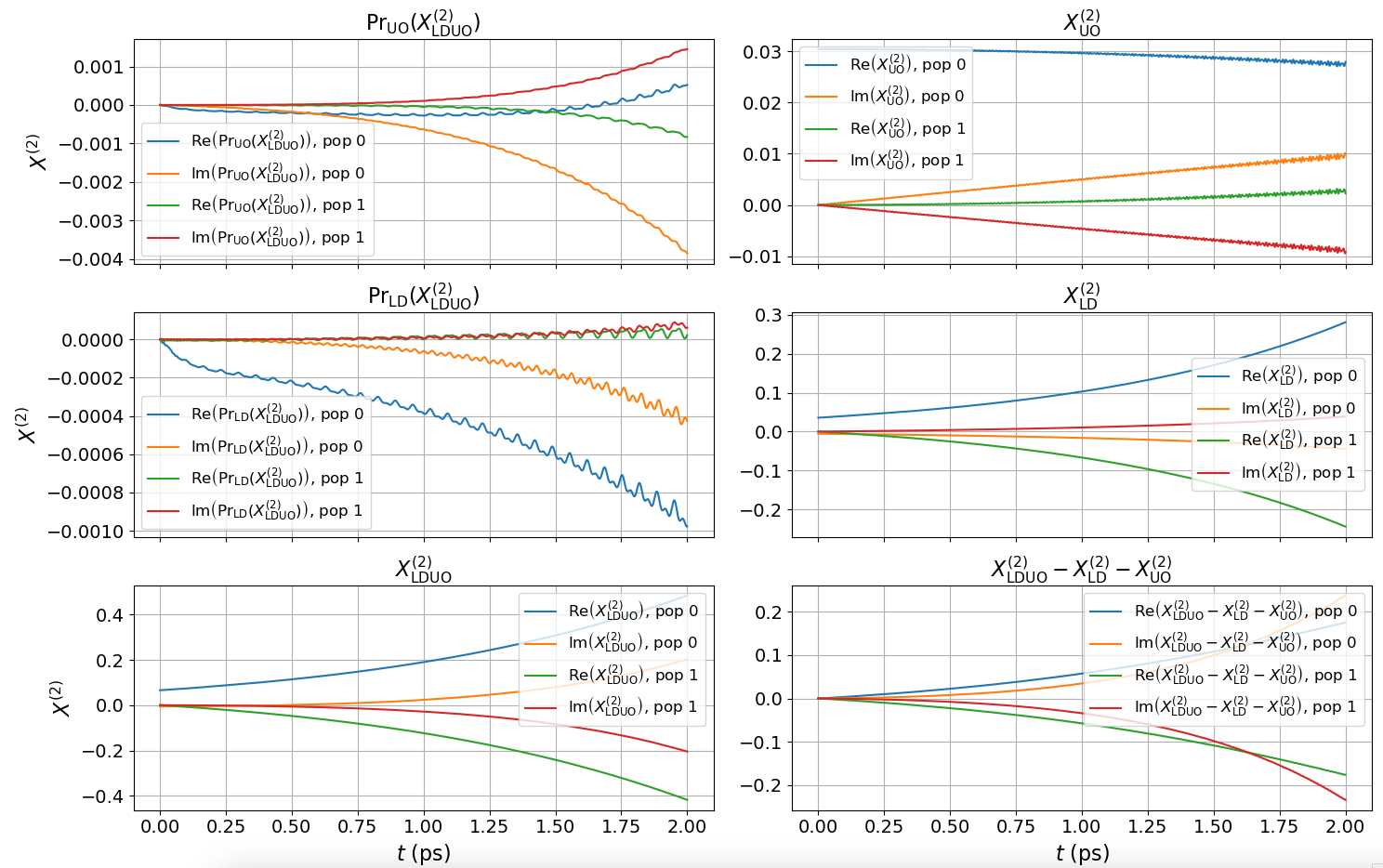}
    \caption{\label{fig:LDUO_diff_2} Tier 2 expectation values. Column 1, the bath coordinate of the full LDUO, projected onto the UO plane, projected onto the LD plane, or unaltered. Column 2, the bath coordinate of the UO, the LD, and the difference between the full bath coordinate and the sum of its parts.}
\end{figure*}


Figure \ref{fig:LDUO_diff_1} presents a decomposition of the bath coordinate expectation values \( X^{(1)}(t) \) for the LDUO model and its isolated constituents: a pure undamped oscillator (UO), a pure overdamped Lorentz–Drude bath (LD), and a canonical two-bath model (LDUO). These results are treated as first order moments which present the mean coordinate of the bath. 

In the top left panel, the projected undamped component of the LDUO model is shown. Here we see an oscillatory structure with the amplitude of the expectation values gradually growing over the lifetime of the trajectory consistent with an increasing mean bath coordinate. This growth appears to arise from the coherent accumulation of energy via the system from the indirect bath-system-bath coupling in the LDUO model. The undamped mode lacks a dissipation channel and hence no equilibrium can be reached, in this projected subspace and the bath continues to build up displacement as it mirrors the system’s dynamics. We further note that, while equilibrium might not be reached with a subspace of the Matsubara volume, equilibrium would eventually be reached by the total system overall. The top right panel shows the isolated UO response, that is, a trajectory with a single undamped bath.  The bath dynamics are governed by a pair of purely imaginary frequencies \( \pm i\omega_{\text{UO}} \), corresponding to Matsubara frequencies arising from the undamped spectral component. These modes correspond to coherent, undamped oscillations in the bath coordinate, which persist indefinitely and maintain constant amplitude and a mean bounded between $0$ and $0.004$. We note that there is no obvious growth, unlike the left-hand side, because there is no other bath from which energy and information can be transferred and as such the mean of the bath coordinate is oscillatory rather than growing. This reflects the fundamentally energy-conserving nature of the undamped mode. Despite being part of an open system, the underdamped bath and system molecule effectively forms a closed system where energy and information is continually exchanging between the two without any loss.

The middle panels display the response of the overdamped bath component. The single bath LD model (middle right) shows purely monotonic decay of entirely real \( X^{(1)} \), consistent with the lack of internal coherence or oscillatory structure. That is, the bath acts as a simple sink, rapidly losing memory of the system through monotonic loss of information. The projected LD component from the LDUO model (middle left) behaves similarly, although modified slightly by its indirect coupling to the UO component. Crucially, there is now an imaginary component which appears because of the system-bath-system coupling that is present in the LDUO model. Additionally, these behaviors underscore the difference between dissipative and conservative dynamics: the LD component loses memory and this facilitates equilibration, while the UO component retains memory. Interestingly this appears to be amplified in the case of the LDUO, where the baths are indirectly coupled via the system.

The bath coordinate of the full LDUO model (bottom left) reveals a hybrid picture, where the bath coordinate contains both persistent oscillations and decaying components. To test that we are definitely seeing information exchange between the baths, the bottom right panel shows a difference plot for \( X^{(1)} \), with contributions from the individual baths subtracted from that of the total LDUO. We see that the latter is not simply the sum of UO and LD parts. That is, nontrivial residuals remain due to interaction between the coherent and dissipative channels. It is therefore clear that strong bath-system-bath coupling occurs within the LDUO model.  

Figure \ref{fig:LDUO_diff_2} shows the results for a decomposition of the bath coordinate expectation values as a second order moment \( X^{(2)}(t) \). These results can be interpreted in terms of two phonon processes as variances of the bath coordinate from the total Matsubara modes present in $A_0$. 

In the top left panel, the projected undamped component of the full system, shows a slow growth in curvature for both the real and imaginary parts of \( X^{(2)} \). This reflects the coherent and persistent build-up of fluctuations in the undamped subspace, driven by significant memory effects. Since no dissipation pathway exists for the undamped mode, the bath coordinate is displaced over time as reflected in the increased bath variance, as a result of the bath-system-bath coupling. In contrast, the top right panel shows the single bath UO trajectory. Unlike the projected LDUO component, the amplitude of the real parts of \( X^{(2)} \) grows symmetrically for the two populations. This indicates evolution of the isolated bath coordinate, coherently oscillating under purely imaginary Matsubara-like frequencies \( \pm i \omega_{\text{UO}} \). Consequently, the variance of the imaginary component increases linearly due to the constant linear driving of the system at these Matsubara frequencies. As there is no dissipative channel or external energy source, the isolated UO response remains bounded and symmetric. 

The middle panels show the corresponding results for the overdamped bath component. The projected LD component from the two bath model (middle left panel) exhibits an oscillatory but secular decay in the real part of \( X^{(2)} \) population zero, and a slight growth for population one. The oscillatory behaviour displays a striking difference compared with the middle right panel, where while there is still a secular divergence in population zero, there is also a much more pronounced growth in population one. Additionally, because there are no oscillatory features present within the signals, there is a much more symmetric variance between the populations. This is consistent with a bath that acts purely as a memoryless sink as the system continuously loses information to the bath. These panels, once again highlight the strong sensitivity to bath coordinate displacements introduced via indirect system-bath-system coupling in the LDUO model.

In the bottom left panel, the full \( X^{(2)} \) from LDUO model shows that both real and imaginary parts diverge over time, suggesting strong population-dependent dressing of the bath with long-lived vibrational and thermal contributions. To examine whether the LDUO result can be decomposed additively, the bottom right panel plots the difference \( X^{(2)}_{\mathrm{LDUO}} - X^{(2)}_{\mathrm{LD}} - X^{(2)}_{\mathrm{UO}} \). Substantial non-zero residuals remain, confirming that the LDUO result is not simply the sum of independent underdamped and overdamped baths. This confirms the presence of significant bath-bath correlations mediated via the system, and highlights that the fluctuation dynamics of the environment in LDUO cannot be interpreted as a linear superposition of simpler baths.

\section{Conclusion}

The derivation of our LDUO model was motivated by the desire to remove superfluious canonical damping from our BVM model. We indeed show that we get qualitatively similar spectra to those of the uHEOM BVM while dramatically improving the computational efficiency, reducing the CPU time by $99.4\%$, or more, in the weak coupling case. 

Perhaps more interestingly,  the development of this model has also demonstrated that there is a strong bath-system-bath coupling present in two bath models which is not present in single bath alternatives, via first and second order bath coordinate expectation values. The bath-system-bath coupling, inherent in this model leads to indirect communication (information transfer) between the baths. Practically, such nonlinear couplings can have a profound impact on system dynamics: for example in a system with electronic dephasing and vibrational relaxation baths the bath-system-bath coupling induces non-trivial electronic-vibrational coupling from the dephasing and vibrational processes which would be absent in a single bath approach. Future work could look at the effect of combining baths via tensor product on bath dynamics.

\begin{acknowledgements}
    The research presented in this paper was carried out on the High Performance Computing Cluster supported by the Research and Specialist Computing Support service at the University of East Anglia. B.S.H thanks the Faculty of Science, University of East Anglia for studentship funding. G.A.J. and D.G. acknowledge support from the Engineering and Physical Sciences Research Council under Awards No. EP/V00817X/1. B.S.H thanks Dr. Joachim Seibt for valuable discussion on the HEOM. 
\end{acknowledgements}

\section*{Conflicts of interest}
The authors have no conflicts of interest to disclose. 

\section*{Data Availability}
The data that support the findings of this contribution are available from the corresponding authors upon reasonable request.

\section*{References}
\bibliography{LDUO_bib.bib}

\end{document}